# Case Studies: Effective Approaches for Navigating Cross-Border Cloud Data Transfers Amid U.S. Government Privacy and Safety Concerns

**Motunrayo Adebayo**

## ABSTRACT

This study attempts to explain the impact of information exchange from one country to another, as well as the legal and technological implications for these exchanges. Due to the emergence of cloud technology, possibilities for free exchange of information between countries have increased rapidly, as it has become possible to save information in a country and access it in almost any part of the world. Countries all around the world have been confronted with developing frameworks to facilitate this process, although there are significant challenges which must be confronted on legal and technological fronts, as loopholes in the framework adopted by countries may hinder free access to information stored on cloud, and also compromise data privacy. Cloud technology is impacting a lot of issues, including domestic and international businesses, hence the need for a study to propose measures for safe exchange of information using cloud technology.

## INTRODUCTION

Evolution has been one of the major elements in the world of information technology, as new developments emerge on a regular basis. One of the major improvements in cloud computing is that computational resources under it can be easily shared by many simultaneous remote users and can be scaled up or down with demand (Herbst et al., 2013). This could lead to a significant reduction in operational costs and increase the ease of service providers and users. For example, private cloud systems can be developed to meet the evolving IT system requirements of consumers. Public users can store their audio, video, photos, and documents online instead of personal computers at home. This gives users the freedom to access their documents wherever they can find the means to access the Internet. These upgrading possibilities offer substantial business opportunities, but have potential legal difficulties for personal data protection, particularly when it comes to exchanging information between countries.

A key feature of cloud technology is that it is transnational in nature. Cloud technology allows data transmissions that can run across the world. Data processing requirements vary from one country to another, in terms of load capacity, time of day, and other factors. These decisions may even be made by machines rather than human beings sometimes (Schwartz, 2013). Therefore, cloud users may not be able to know the true location of physical infrastructure as well as the actual location of personal data. Although traditional Internet technology allows cross border data transactions, in these transactions, data owners and processors seem to have better control over the data they process. However, cloud technology has arguably increased the variableness of data control, as well as the uncertainty of compliance with legal requirements. This is particularly pertinent in the current international trade environment. A company, either operating internationally or dealing with international clients, must comply with the laws in relevant countries. If the company uses a cloud service provided by any third party, the provider is required to ensure compliance with the relevant laws of the country in question. For example, if an Australian company has entered into an agreement with a cloud provider in the United States, but the cloud provider hosts the data on a server in the EU, this means that the company needs to comply with laws in Australia, the U.S., and the EU.

The exchange of information is very critical in the world, as the global economy has gone digital, and digital operations are affecting businesses including local and international trade. In today's world, businesses rely on the exchange of data across borders. Accordingly, data has come to the center of countries' regulatory concerns. The Information Technology and Innovation Foundation (ITIF) in a 2021 report pointed out that the number of data-localization measures that are currently in operation has doubled between 2017 and 2021. In 2017, only 35 countries implemented 67 restrictive measures, while the number rose to 62 countries implementing 144 measures, with more under consideration as at 2021 (ITIF, 2021).

The 'Trusted Cloud Principles', an agreement to protect client data privacy and security regardless of local boundaries, were recently developed by top cloud service providers. One of these fundamental tenets calls for the avoidance of data residency laws and underlines the need for government support for the international interchange of data as a driver of innovation, efficiency, and security. National borders are less clear in the linked digital world of today. It is commonly acknowledged that allowing cross-border data sharing promotes economic growth and internet trade. Global governments must adopt laws that protect people's privacy and personal information while still promoting cross-border data flow. Regrettably, the issue of inconsistent and divergent legislation between nations is getting worse, resulting in requests for local data storage and limits on data movement.

The interchange of digital services across borders and global data flows, however, has significantly increased in recent years. One gigabit per person per day, or three zettabytes, of internet traffic was generated globally in 2020, according to World Bank research. It is anticipated that this enormous volume of data will double soon, spurring an increase in global trade. By facilitating the interchange of commodities, enhancing productivity, and cutting costs, cross-border data flows are essential to commerce. Additionally, they are necessary for carrying out transactions in digital services. Data transmission has been a major factor in the exponential rise of international trade, demonstrating the interdependent link between cross-border data flows and global trade. In fact, it is difficult to think of a global commerce transaction that does not entail data transfer.

A nation's economic prosperity depends on having a well-designed legal framework for the transmission of data across international boundaries. Building a strong system should be a top priority considering the expanding global data exchanges and possible hazards such threats to national security, data breaches, and privacy concerns. By ensuring that personal data is protected throughout transfers and preventing misuse or exploitation, this framework strives to achieve its goal.

Cross-border data transfers are now made possible by several mechanisms, including the Privacy Shield Framework between the US and the EU, the APEC Privacy Framework, and the General Data Protection Regulations (GDPR) in the EU. Regardless of where they are located, firms processing the personal data of EU people must abide by the GDPR, which is a Data governance is also at the center of the US-China technology conflict. On the one hand, the US and China are competing to be the rule-maker in the digital age, with access to data as a decisive and deployable instrument in enhancing firms' strategy and competitiveness. On the other hand, the position of the two countries represents the conflict between two ideologies about data governance. While China adopts a more restrictive approach on cross-border data transfer, the US has supported free flow of data. However, in a significant twist, China has shown signs of easing control of data transfers since the second half of 2023, while the US has taken a step back from its full support of free data transfers.

## THE EVOLVING LANDSCAPE OF DATA TRANSFER REGULATIONS

China is among the countries believed to have the most restrictive administrations on cross border data transfers in the world. In 2021, the Chinese congress passed the Personal Information Protection Law, which specified the conditions for companies that need to transfer large amount of data out of the country. The government also issued more detailed regulations on the procedure, which were supposed to be inimical to foreign investors and transnational enterprises, due to the walls to technology transfer and transnational trade. The Personal Information Protection Law's compliance mechanisms represent a restrictive approach to data governance, like the GDPR used by the European Union. The GDPR offers analogous assessment mechanisms that permit transfer of data outside Europe, if it's certain that those areas offer sufficient position of data protection. In discrepancy, the US submitted an offer to the World Trade Organization (WTO) in 2019 to guarantee free data transfers and bar forced data localization in member countries. These conflicts of testaments among the two of the foremost technologically advanced countries demonstrate a dicker between digital trade and cybersecurity enterprises.

In a different twist, in October 2023, the United States Trade Representative (USTR) Katherine Tai blazoned that it had dropped these digital trade demands to give steeper regulations for data transfer control. In discrepancy, on September 28, 2023, China also moved in the contrary direction, proposed to water down some of its CBDT controls by promoting cross-border data flows. The draft exempted:
  a. Data exported as part of international trade, academic cooperation, or cross-border manufacturing and marketing and does not contain "important personal information".
  b. Personal information necessary for the performance of a contract (e.g. cross-border transactions, flight and hotel reservations, visa applications)
  c. Employee information necessary for human resource management
  d. Personal information necessary to protect the life, health, and property safety of a natural person in an emergency
  e. Organizations that export less than 10,000 individuals' personal information within one year, from the PIPL's CBDT requirements.

The ideal of both countries' moves indicates the binary effect of cross-border data flows. Data localization is generally justified on the base of cybersecurity, particular sequestration, and digital sovereignty but is frequently attacked for impeding transnational trade, yielding digital protectionism, and blocking the creation of an open internet.

**Data Protection Regulations and Compliance**
With the preface of important legislation like the GDPR in the European Union, transnational data protection laws have been made tough. The GDPR imposes tight guidelines that limit the gathering, processing, and transfer of particular data belonging to EU citizens, outside the borders of the mainland. The rules apply, anyhow of the position of processing. The California Consumer sequestration Act (CCPA) is notable as innovative law in the United States. CCPA, which has its roots in California, gives consumers enormous control over their particular information and thereby empowers them. The sequestration of guests is enhanced by strict data protection rules that give them more control over their particular information. Data breaches and sequestration dishonors have raised client mindfulness of the value of their data, leading to more careful online conduct. Consumers are more likely to engage in online conditioning and deals when they're certain that companies are clinging to strict data sequestration norms.

**International Trade and Foreign Investment**
International Trade and Foreign Investment Opponents of data localization generally emphasize that a restrictive approach has ineffective goods and increases enterprises' compliance costs. This is the primary reason why the US favors an unrestricted approach that permits free cross-border data flows. The 2019 WTO submission stressed cross-border data inflow as "the lifeblood of transnational trade," citing McKinsey Global Institute that indicated that cross-border data flows generated$ 2.8 trillion in profitable value in 2014 — a lesser impact on world GDP than global trade in goods. China's relaxation of CBDT control in 2023 also signals its desire to revitalize foreign investment. Before the proposed relaxation, foreign investors and MNEs subject to the PIPL's governance were faced with only three options to insure compliance to conduct the assessment, instrument, or recordation procedures which dodge significant costs; to make or cooperate with original data centers to store the information within the border; or to simply exit the request. The release of CBDT control in September 2023 is among the measures espoused by Chinese policymakers to win back overseas investors, as the former months of 2023 had witnessed a mass capital outpour in the country. Meanwhile, the draft vittles also allow free trade areas to legislate and apply separate measures to grease transnational trade.

### CHALLENGES AFFECTING CROSS-BORDER CLOUD DATA TRANSFERS
Tian (2017) identified 3 main challenges for cross-border cloud data transfers. He outlined them as privacy challenges, jurisdiction challenges, and convergence challenges. Jurisdiction refers to legal clauses contained in service level agreements (SLA), as well as the impact of government intervention in data sharing and data location. Travis (2016) contends that when a cloud service provider has data centers in various countries, in addition to data breach and abuse by an individual or any third-party overseas, there is a risk of data breach by governments, by both local and foreign régimes. Governments in many countries have regulations to compel cloud service providers to grant governments entry into personal data in certain circumstances, such as matters concerning national security or law enforcement, which may be difficult for cloud service providers to reject (USA Patriot Act, 2003).

On the other hand, privacy challenges encompass data transfer, and legal compliance issues. The globalized nature of data transfer in contrast with the limitations of national law has created a patchwork system of laws that apply at the domestic level, although the storage and transfer of data is international. As such, when cloud service providers set up data storage centers, to avoid legal difficulties they need to be aware of and adhere to the laws in the country in which their data storage centers are located, particularly the laws that affect cross-border data transfer. Likewise, when cloud users choose their cloud service providers, they ought to be informed of the location of their cloud service provider's data storage infrastructure, and the potential risk of their personal data being unprotected by the laws of the country where the user is based, which may result in a false expectation of privacy (Tian, 2017).

Tian (2017) further posits that convergence challenges for cross-border personal data protection jreflect two areas, which are: challenges from the convergence of technology, challenges from the convergence of laws. He emphasizes the need for any cloud service providers, users, and regulators, particularly those who must deal with cross-country personal data exchanges, to carefully consider the three major challenges which have been clarified. These three challenges are deeply interlinked, and certain overlaps may exist among the three. The overlaps and interactions between the three challenges provide evidence for and reflect upon the nature of integration and convergence in the information technology industry.

## SUGGESTED APPROACHES FOR SAFE CROSS-BORDER CLOUD DATA TRANSFERS

There have been several challenges associated with transfer of data across borders, including evolving regulations, variance in technological infrastructure, data security, and differing data protection laws. The following encompasses approaches which must be considered in administering safe cross border data transfers.

**Rule of Law**
The foundation for a trust-based framework for cross-border data flows is based upon the participating countries sharing demonstrable commitment to democratic governance under the rule of law. With that foundation, countries can have confidence that legal obligations to protect rights will be respected and enforced. A democracy governed by rule of law ensures "political rights, civil liberties, and mechanisms of accountability which in turn affirm the political equality of all citizens and constrain potential abuses of state power" O'Donnell (2004). In authoritarian regimes, "power is concentrated in the hands of a single leader or small elite", and the regime governs without the consent of its citizens (Lindstaedt, 2023). Under authoritarianism, there are no legitimate accountability mechanisms, and transfer of executive power does not exist. Countries seeking to benefit from the framework should meet internationally recognized criteria for democratic governance under the rule of law (UN, 2016). If a receiving country does not meet those criteria, then sending countries may well need to follow individualized approaches to restrict data flows and ensure rights are protected.

**Rights-Protective**
Countries committed to the rule of law prioritize individual rights and strive to ensure that the rights granted to their citizens or residents are upheld in international data transfers. Therefore, any framework for these data flows must incorporate effective safeguards that genuinely protect individual rights and prevent abuse or misuse of data by both private-sector entities and governments. Such safeguards should guard against access that contradicts democratic values and the rule of law, as well as any access that is unreasonable, arbitrary, or disproportionate. A rights-protective framework should also include accountability measures to ensure that data processors are properly implementing safeguards, which should encompass both internal and external oversight, along with avenues for individual redress. In summary, the framework must ensure that processing entities—whether governmental or private—respect individuals' privacy and fundamental rights in the recipient country in a manner that aligns with, but is not identical to, practices in the originating country.

**Practicable/Adaptable**
The framework should recognize that countries have distinct legal systems, allowing each to develop its own safeguards and accountability measures. Recipient countries shouldn't be required to overhaul their legal structures to mimic the laws of the originating country or simply accept those already in place elsewhere. However, countries must not become complacent in their democratic governance; they should address any legal or procedural gaps and enhance deficiencies to ensure meaningful safeguards and effective accountability mechanisms (Docksey, 2019).

**Scalable**
The framework needs to keep pace with the rapid, expansive, and global nature of international data flows, facilitating fair and efficient determinations regarding cross-border data transfers based on agreed-upon, objective criteria. Existing building blocks for such a framework include efforts on cross-border transfer mechanisms under the GDPR and similar regulations outside the EU, as well as the EU-US Data Privacy Framework (EU, 2023) and the OECD Privacy Guidelines related to the Asia-Pacific Economic Cooperation (APEC), among others.

## CONCLUSIONS
The strategic significance of data is widely recognized in the digital era, yet regulatory approaches differ significantly worldwide. The US adopts a more unrestricted view of data sharing, treating data as property and promoting its free flow across borders. In contrast, the EU and China take a more restrictive stance, leading to a data localization effect. This regulatory divergence stems from differing philosophical views on data and personal property. The US's approach aligns with its liberal market capitalism ideology, while Europe champions data protection through the GDPR, emphasizing personal data rights as fundamental human rights. China, prioritizing personal privacy, upholds cyber sovereignty, ensuring state access to all data generated within its borders. These core differences highlight the rationale behind varying legal frameworks for data transfers.

The global divergence in data regulation increases costs for multinational enterprises and complicates domestic law enforcement efforts. Significant expenses arise from establishing data storage facilities, conducting necessary reviews, and implementing related measures, all of which elevate compliance costs for large IT companies managing cross-border data transfers. Even with such measures in place, substantial compliance risks remain due to conflicting regulatory requirements across countries. For instance, China's PIPL has complicated legal proceedings in US federal courts, leaving the legal aspects of data transfers largely unresolved.

Courts play a crucial role in addressing these conflicts by applying established legal principles and balancing the interests of the parties involved. Their decisions carry significant implications for international business and data flows, impacting foreign trade relations.

Despite these fundamental differences, signs of convergence appeared in 2023 regarding countries' attitudes toward cross-border data transfers (CBDT). In September 2023, China's CAC released a draft policy paper proposing exemptions from mandatory assessments for certain transfers, while in October, the Office of the USTR dropped its demands at the WTO aimed at promoting free cross-border data flows among member states.

Navigating the complexities of data transfers has profound implications for large IT companies and the future of international data governance. As the US and China explore these uncharted territories, there is potential for collaboration. Aligning data collection, storage, transfer, and consent regulations could foster smoother relations. Thus, the interplay between data privacy, cross-border litigation, and technology's intersection with the rule of law will continue to influence our world in the years ahead.

## References

George Yijun Tian. Current Issues of Cross-Border Personal Data Protection in The Context of Cloud Computing and Trans-Pacific Partnership Agreement: Join or Withdraw. Wisconsin International Law Journal, Vol 34, No 2, pp 367 – 408 (2017).

Guillermo O'Donnell. The Quality of Democracy: Why the Rule of Law Matters, Journal of Democracy, Vol. 15, No. 4, 32 (2004).

Nikolas Roman Herbst, Samuel Kounev & Ralf Reussner. Elasticity in Cloud Computing: What It Is, And What It Is Not (2013).

Paul M. Schwartz, EU Privacy and the Cloud: Consent and Jurisdiction Under the Proposed Regulation, 12 BNA PRIVACY AND SECURITY L. REP. 718, 718 (2013)

Natasha Lindstaedt. Authoritarianism, Encyclopedia Britannica, June 2023.
https://www.britannica.com/topic/authoritarianism

United Nations. What is the Rule of Law. https://www.un.org/ruleoflaw/what-is-the-rule-of-law/; Summit for Democracy, Joint Statement and Call to Action on The Rule of Law and People-Centered Justice: Renewing a Core Pillar of Democracy, USAID. https://www.usaid.gov/sites/default/files/2023-04/Joint-Statement-Call-to-Action-onthe- Rule-of-Law-and-PCJ-April-14-2023.pdf; Venice Commission of the Council of Europe, The Rule of Law Checklist, March 2016.
https://www.venice.coe.int/images/SITE%20IMAGES/Publications/Rule_of_Law_Check_List.pdf

Christopher Docksey. Keynote on Accountability At the 41st Conference of Data Protection and Privacy Commissioners in Tirana, Albania, Oct. 31, 2019. https://informationaccountability.org/2019/10/christopher-

docksey-keynote-on-accountability-at-the-41st-conference-of-data-protection-and-privacy-commissioners-24-october-2019-in-tirana-albania/

Christopher Docksey. Article 24 in C. Kuner, L. A. Bygrave, and C. Docksey (eds.), The EU General Data Protection Regulation: A Commentary, OUP, 2020, pp. 555–570.

OECD Privacy Guidelines, 1980.https://legalinstruments.oecd.org/en/instruments/OECD-LEGAL-0188.

Asia-Pacific Economic Cooperation (APEC), APEC Privacy Framework, Dec. 2005. https://www.apec.org/publications/2005/12/apec-privacy-framework.

US Department of Commerce, Global Cross-Border Privacy Rules Declaration, 2022. https://www.commerce.gov/global-cross-border-privacy-rules-declaration.

Alan Travis, UK Security Agencies Unlawfully Collected Data for 17 Years, Court Rules, The Guardian, Oct. 18, 2016 in Robert McMillan & Jennifer Valentino-Devries, Russian Hacks Show Cybersecurity Limits, Wall Street Journal (2016).

USA Patriot Act of 2001, 18 U.S.C. 2511(2) (2006); Allen & Overy, The EU General Data Protection Regulation (2016)
http://www.allenovery.com/SiteCollectionDocuments/Radical%20changes%20to%20European %20data%20protection%20legislation.pdf.